%% file: example_paper.tex
\documentclass{article}
\usepackage{subcaption}
\usepackage{microtype}
\usepackage{graphicx}
\usepackage{booktabs} 
\usepackage{multirow}

\usepackage{hyperref}

\usepackage[accepted]{icml2025}

\usepackage{amsmath}
\usepackage{amssymb}
\usepackage{mathtools}
\usepackage{amsthm}
\usepackage{dsfont}
\usepackage{tcolorbox}
\usepackage{enumitem}

\usepackage[capitalize,noabbrev]{cleveref}

\theoremstyle{plain}

\theoremstyle{definition}

\theoremstyle{remark}

\usepackage[textsize=tiny]{todonotes}

\icmltitlerunning{Learning to Solve and Verify}

\begin{document}

\twocolumn[
\icmltitle{Learning to Solve and Verify:\\
           A Self-Play Framework for Code and Test Generation}



\icmlsetsymbol{equal}{*}

\begin{icmlauthorlist}
\icmlauthor{Zi Lin}{equal,yyy}
\icmlauthor{Sheng Shen}{comp}
\icmlauthor{Ilia Kulikov}{comp}
\icmlauthor{Jingbo Shang}{yyy}
\icmlauthor{Jason Weston}{comp}
\icmlauthor{Yixin Nie}{comp}\\
\icmlauthor{\textbf{Meta}}{comp}
\icmlauthor{\textbf{UCSD}}{yyy}

\end{icmlauthorlist}

\icmlaffiliation{yyy}{University of California, San Diego}
\icmlaffiliation{comp}{Meta}

\icmlcorrespondingauthor{Zi Lin}{lzi@ucsd.edu}
\icmlcorrespondingauthor{Yixin Nie}{ynie@meta.com}

\icmlkeywords{Machine Learning, ICML}

\vskip 0.3in
]



\printAffiliationsAndNotice{*Work done during internship at Meta GenAI.\ } 

\newcommand{\jingbo}[1]{\textcolor{blue}{\textbf{Jingbo:} #1}}

\begin{abstract}
Recent advances in large language models (LLMs) have improved their performance on coding benchmarks. 
However, improvement is plateauing due to the exhaustion of readily available high-quality data.
Prior work has shown the potential of synthetic \textit{self-instruct} data, but naively training on a model’s own outputs can cause error accumulation, especially in coding tasks, where generalization may collapse due to overly simple or erroneous training data, highlighting the need for rigorous quality checks on synthetic data.
In this work, we explore an effective approach whereby the model itself verifies the correctness of 
its own data. 
We thus propose {\sc Sol-Ver}, a self-play solver-verifier framework that jointly improves a single model’s code and test generation capacity. By iteratively refining code (LLM-as-a-solver) and tests (LLM-as-a-verifier) together, we boost both capabilities without relying on human annotations or larger teacher models.
%
Experiments with the Llama 3.1 8B model demonstrate substantial performance enhancements, achieving average relative improvements of 19.63\% in code generation and 17.49\% in test generation on MBPP and LiveCodeBench.
\end{abstract}

\input{sections/01.introduction}
\input{sections/02.related-work}
\input{sections/03.methods}
\input{sections/04.experiments}
\input{sections/05.ablation-study}
\input{sections/06.conclusion}

\bibliography{example_paper}
\bibliographystyle{icml2025}

\newpage
\appendix
\onecolumn
\input{sections/appendix}



\end{document}

%% file: sections/01.introduction.tex
\section{Introduction}

Large language models (LLMs) have demonstrated impressive ability in code generation, significantly enhancing the programming efficiency and productivity of human developers~\cite{li2022competition,roziere2023code,codealpaca}.
The ability to code is largely due to high-quality online coding resources, e.g., coding problem and human-rewritten solutions.
However, as these supervised data sources saturate, LLM improvement is diminishing, with data scarcity becoming a key bottleneck for further progress.

To address scarce supervised data for code generation, recent studies use synthetic data techniques like {\sc Self-Instruct}~\cite{wang2023self} to augmenting LLM training sets. 
Typically, a high-capacity teacher LLM generates code responses to designed instructions, and this data is then employed to fine-tune a student LLM, thereby enhancing its code generation abilities.
Although synthetic code data produced in this manner has demonstrated success, it relies on the availability of a strong teacher model, presumably with a larger parameter size and higher computation costs.
Additionally, existing work has shown that training a model on data generated by itself is ineffective because errors introduced during generation tend to accumulate over iterations~\cite{dubey2024llama}. As a result, there is a critical need for effective methods to verify the generated data.

While evaluating generated code correctness is challenging, often requiring expert human intervention, recent LLM-as-a-judge efforts involve models executing generated code against self-generated unit tests~\cite{mcaleese2024llm,alshahwan2024automated,dong2024self,chen2024b4,chen2022codet,chen2023selfdebug,dubey2024llama}. However, a critical bottleneck emerges: an LLM's proficiency as a \textit{verifier} (generating effective unit tests) significantly lags its capability as a \textit{solver} (generating code solutions), a disparity we quantify in Section~\ref{sec:base-performance}). This gap is largely attributable to the scarcity of high-quality, diverse unit test generation data used during LLM fine-tuning, which predominantly focuses only on code generation.

To address this imbalance and unlock a new avenue for data generation, we introduce \textsc{Sol-Ver}, a self-play solver-verifier framework to iteratively train a model for both code and test generation. The main idea is to let the LLM-as-a-solver and LLM-as-a-verifier help each other. Specifically, we ask the model to generate code solutions and unit tests for the same set of coding problems. By  executing the generated test against the generated code, we obtain feedback for training, involving two steps: (1) SFT training: we take the passed examples for fine-tuning the model, and (2) DPO training: we take both passed and failed examples as preference pairs to further train the model aligning with the preference. These training steps are for both code generation and unit test generation, and they can be repeated in an iterative manner. 

The experimental results on Llama 3.1 8B model show that we can successfully improve the model’s performance on both code and test generation without relying on human-annotated data or larger models. Specifically, on MBPP and LiveCodeBench, we achieve an average of 19.63\% and 17.49\% relative improvement for code and test generation respectively.

\begin{figure*}[h]
    \centering
    \includegraphics[width=0.95\textwidth]{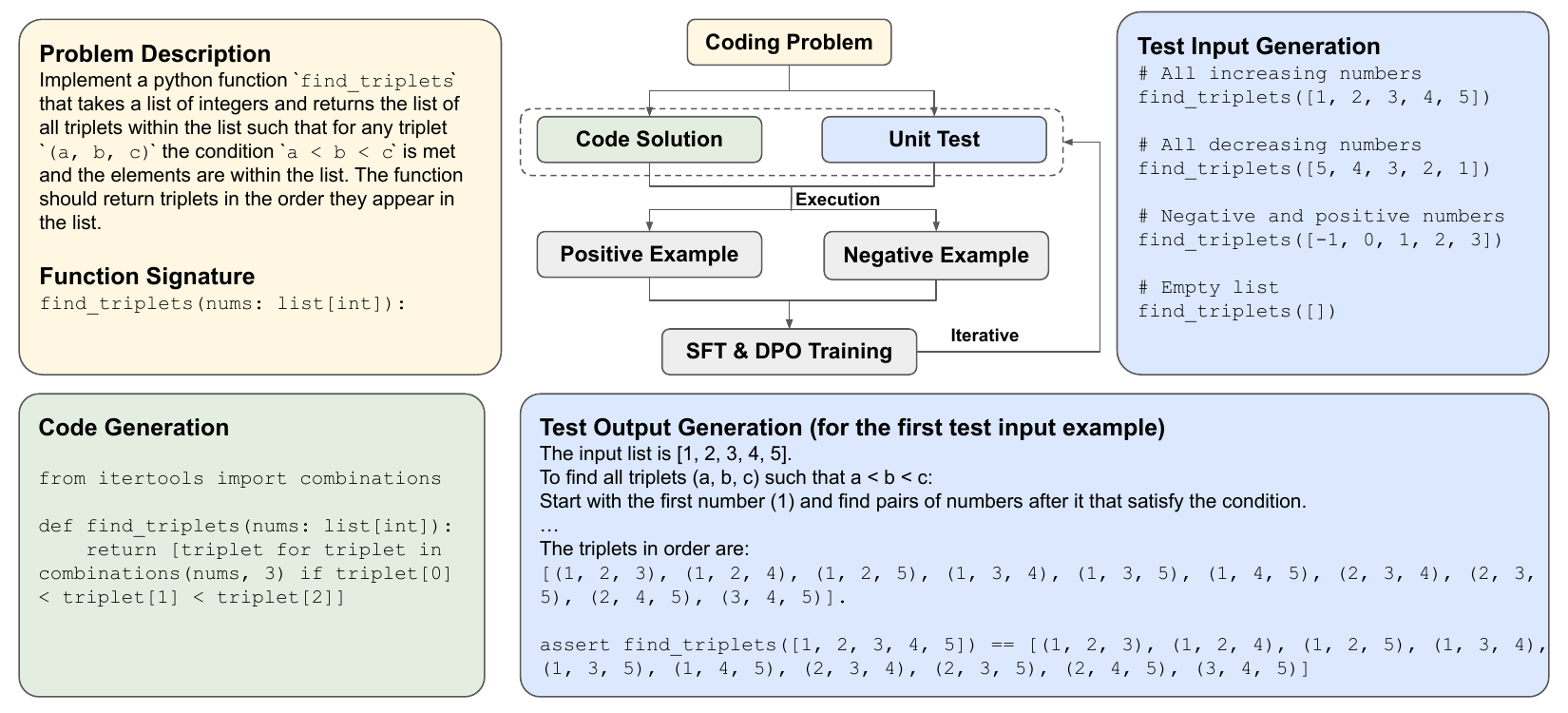}
    \caption{An overview of the {\sc Sol-Ver} framework.
We train an LLM to both generate coding solutions (solver) and unit tests (verifier) in an iterative self-play framework, whereby synthetic preference pairs are constructed at each iteration depending on whether the code passes the generated tests or not.
We show that this approach enables the model to self-improve in both capabilities (see \autoref{tab:full-performance}).
    } 
    \label{fig:enter-label}
\end{figure*}

In summary, our work makes the following contributions:
\begin{itemize}[nosep, leftmargin=*]
    \item \textbf{Identification of a {\em critical gap}:} We empirically demonstrate and analyze the significant gap in LLMs' abilities between code generation and unit test generation, motivating the need for targeted improvements in test generation.
    \item \textbf{Novel \textit{Self-Play} Framework:} We propose a novel iterative framework where the model simultaneously functions as a code \textit{solver} and a \textit{verifier}. This methodology effectively self-aligns the model's outputs with desired performance criteria without relying on external annotations or teacher models.
    \item \textbf{High-Quality Synthetic Data Generation:} We contribute a generalizable method for creating high-quality synthetic data for both code and unit test generation. This data augmentation approach can be extended to various model training scenarios in the coding domain.
\end{itemize}

%% file: sections/02.related-work.tex
\section{Related Work}
Previous studies applying scaling laws to the training of foundation models have highlighted the critical role of the data size~\cite{kaplan2020scaling,hoffmann2022training,chung2022scaling}. To address the need for larger datasets, synthetic data generation has become a popular and cost-effective solution, which leverages advanced LLMs to produce high-quality data.
One notable method is {\sc Self-Instruct}~\cite{wang2023self,Taori2023Alpaca}, which employs a pre-trained LLM to generate instruction-output pairs from a small seed dataset.

For code generation, previous work devises synthetic coding instructions using a stronger teacher model (e.g., ChatGPT or GPT-4) and then finetunes a weaker student model (e.g., {\sc CodeAlpaca}~\cite{codealpaca} and {\sc CodeLlama}~\cite{roziere2023code}) with the generated data to distill knowledge from the teacher. For example, code alpaca consists of 20K automatically generated code instructions by applying {\sc Self-Instruct} on ChatGPT using 21 seed tasks. To enhance the code abilities of LLMs, \citet{luo2023wizardcoder} proposes \textit{Code Eval-Instruct} that employs heuristics to increase the complexity of seed code instructions. Magicoder~\cite{wei2024magicoder} proposes to generate coding problems by drawing inspiration from snippets collected from open source code.

While previous work has shown significant improvements for models trained on data generated by larger, more competent models, training an LLM on its own generated data is not helpful and can even degrade performance~\cite{zhou2024lima,alemohammad2023thecurseofrecursion}. To prevent the model from learning errors present in its own generated data, some post-processing steps are essential. For example, 
CodeT~\cite{chen2022codet} leverages pre-trained language models to automatically generate test cases for multiple code samples. It executes the generated code against these test cases and employs a dual execution agreement strategy, ensuring both output consistency and agreement among multiple code solutions.
Self-Debug~\cite{chen2023selfdebug} enables models to autonomously identify and fix bugs in their generated code.
Llama 3.1~\cite{dubey2024llama} utilizes error feedback from execution and adopts an iterative self-correction procedure to revise potential errors.
Reinforcement learning from unit test feedback (RLTF) has also been explored to improve code generation by directly optimizing for test pass rates~\cite{le2022rltf, ni2023teaching}
CodeDPO~\cite{zhang2024codedpo} replaces teacher models with a self-generation-and-validation process that uses a PageRank-like algorithm to rank code snippets by correctness and efficiency, yielding diverse preference optimization data without external resources.
Another related line of work, AutoIF~\cite{li2024autoif}, proposes an iterative framework for generating instruction-following data for general tasks by having an LLM play the roles of instructor and respondent, using another LLM as an evaluator, though it does not specifically focus on code generation or the co-evolution of solver and verifier capabilities.

In this work, we propose leveraging both positive and negative examples generated by the model, treating pairs of passing and failing responses as chosen-rejected pairs for Direct Preference Optimization (DPO)~\cite{rafailov2024direct}. Note that our method is complementary to self-correction and RLTF, rather than orthogonal. By improving the quality of unit tests, our framework enhances the accuracy of unit test execution feedback, and thereby can benefit self-correction and RLTF scenarios as well.

\textbf{{\sc Sol-Ver} Compared to CodeDPO:} While CodeDPO~\cite{zhang2024codedpo} also utilizes self-verification for preference data, its efficacy is inherently limited by the initial, and potentially static, quality of the self-generated tests used for verification. If the verifier component is weak, it cannot reliably distinguish high-quality code, potentially leading to suboptimal preference learning. In contrast, {\sc Sol-Ver}'s self-play mechanism is designed to explicitly address this: it \textbf{simultaneously improves both the code solver and the test verifier}. By iteratively enhancing the verifier's ability to generate more discerning and comprehensive tests, {\sc Sol-Ver} breaks this quality ceiling, thereby mitigating the verifier bottleneck and enabling more robust and effective training for both capabilities.

%% file: sections/03.methods.tex
\section{A Self-play Solver-verifier Framework}
\subsection{Problem Formulation}
\label{sec:problem-formulation}
\paragraph{Setup} We consider that an LLM can play two roles:
\vspace{-2mm}
\begin{itemize}[leftmargin=*]
    \item \textbf{Solver ($S$):} Given a coding problem description $P$, it produces a candidate solution $C$ (e.g., a piece of code).
    \item \textbf{Verifier ($V$):} Given a proposed solution $C$ and the original problem $P$, the verifier tries to produce test cases $\mathbf{T}$\footnote{We use bold Italic to represent a set.} (e.g., a set of inputs and expected outputs) and can catch errors in $C$ if it is incorrect. Essentially, it produces and selects challenging unit tests to determine if the code is correct or not.
\end{itemize}

The objective of the solver is to produce a correct solution $C$ that will pass any tests the verifier can come up with. The objective of the verifier is to produce a set of tests $\mathbf{T}$ that will fail any incorrect solutions and thus distinguish correct solutions from incorrect ones.

Let $p(\mathbf{P})$ be the distribution over problem statements. We can think of having a training set of problems or a domain from which we can sample problems. The sampling strategies we consider are detailed  in Section~\ref{sec:practical-settings}.

The solver $S_{\theta}$ is a model parameterized by $\theta$ that, given a problem $P$, generates a candidate solution $C$: $C \sim S_{\theta}(\cdot | P)$. The verifier $V_{\phi}$ is a model parameterized by $\phi$, given the problem $P$ and a candidate solution $C$, generates a test suite $\mathbf{T}$: $\mathbf{T} \sim V_{\phi}(\cdot | P, C)$.

In practice, the solver and verifier can be the same LLM.

\textbf{Scoring Function}\ \ \ We define a function that executes $C$ on the tests $\mathbf{T}$ as $\text{Score}(C, \mathbf{T}) \in [0, 1]$, which is the fraction of tests passed by solution $C$. A score of 1 means $C$ passes every test $T$; a score of 0 means it failed all tests. Formally,
\begin{align}
    \text{Score}(C, \mathbf{T}) = \mathbb{E}_{T\sim \mathbf{T}}[\mathbb{I}(C(T)=\text{expected\_output}(T))]
\end{align}
where $\mathbb{I}$ is the indicator function, and $C(T)$ means running one single test on code solution $C$.

We sample a set of problems $P \sim p(\mathbf{P})$, generate some candidate solutions $C \sim S_{\theta}(\cdot | P)$, and generate candidate tests $\mathbf{T} \sim V_{\phi}(\cdot | P, C)$. We now have tuple $(P, C, \mathbf{T})$. We consider:

\begin{align}
y = \begin{cases}
1 & \text{if } \text{Score}(C,\mathbf{T}) = 1 \text{ (i.e., passes all tests)} \\
0 & \text{otherwise}.
\end{cases}
\end{align}

We employ two stages of  training to make use of both chosen ($y=1$) and rejected ($y=0$) examples for training the solver and verifier, described as follows:

\textbf{Stage 1: SFT Training}\ \ \ For pairs where $y = 1$, we have a correct solution-test pair. These are high-quality examples that reflect desired behavior, i.e., the solution $C$ solves the problem $P$, and the test suite $\mathbf{T}$ properly validates that the solution is correct. We use $(P, C, \mathbf{T}, y = 1)$ tuples to fine-tune the model directly. The training signal here encourages the model (1) as a solver, to generate similar correct solutions for similar problems, and (2) as a verifier, to produce meaningful tests that confirm correctness. We call this the supervised fine-tuning (SFT) stage, where we optimize for both solver and verifier:

\begin{align}
    \mathcal{L}_{\text{SFT}_\text{solver}}(\theta) = -\mathbb{E}_{(P, C, \mathbf{T}):y=1}[\log S_{\theta}(C|P)] \\
    \mathcal{L}_{\text{SFT}_\text{verifier}}(\phi) = -\mathbb{E}_{(P, C, \mathbf{T}):y=1}[\log V_{\phi}(\mathbf{T}|P, C)]
\end{align}

In practice, both objectives can be trained using a mixture of data consisting of chosen examples for solver and verifier.

\textbf{Stage 2: DPO Training}\ \ 
\label{sec:dpo-training}
We now aim to form pairwise comparisons (preferences) to train both solver and verifier roles more effectively. In practice, we adopted the Direct Preference Optimization (DPO) method, but any preference tuning methods can be used at this stage.

For the solver perspective, for each problem $P$, and each chosen tuple $(P, C^+, \mathbf{T}, y=1)$, we find a rejected tuple $(P, C^-, \mathbf{T}, y=0)$. Following standard DPO training~\cite{rafailov2024direct}, we can formulate our policy objective as:
\begin{align}
    &\mathcal{L}_{\text{DPO}_\text{solver}}(S^*_{\theta}; S_{\theta}) = \nonumber \\ &-\mathbb{E}[\log\sigma(\beta\log \frac{S^*_{\theta}(C^+|P)}{S_{\theta}(C^+|P)} - \beta\log \frac{S^*_{\theta}(C^-|P)}{S_{\theta}(C^-|P)}]
\end{align}

where $\beta$ is the hyperparameter to regulate the strength of weight updates; $S_{\theta}(C|P)$ is the probability that our model (with parameter $\theta$) assigns to generating code solution $C$ given problem $P$.

For the verifier perspective, similarly, for each problem $P$, and each chosen tuple $(P, C, \mathbf{T}^+): y=1$, find a rejected tuple $(P, C, \mathbf{T}^-):y=0$\footnote{This can be achieved by selecting any expected output that is not equal to the chosen one in the sampling space for $T$.}. The verifier-related DPO loss is then:
\begin{align}
    &\mathcal{L}_{\text{DPO}_\text{verifier}}(V^*_{\phi}; V_{\phi}) = \nonumber \\ &-\mathbb{E}[\log\sigma(\beta\log \frac{V^*_{\phi}(\mathbf{T}^+|P, C)}{V_{\phi}(\mathbf{T}^+|P, C)} - \beta\log \frac{V^*_{\phi}(\mathbf{T}^-|P, C)}{V_{\phi}(\mathbf{T}^-|P, C)}]
\end{align}

\subsection{Synthetic Data Generation}
\label{sec:practical-settings}
In this section, we describe our approach to generating the synthetic data, including problem description generation, code generation, test generation and preference data generation. All related prompts can be found in Appendix~\ref{app:prompt-template}.

\textbf{Problem Description Generation}\ \ \ Following Magicoder~\cite{wei2024magicoder}, we generate a large collection of programming problem descriptions that span a diverse range of topics, including those in the long tail distribution. To achieve this diversity, we sample random code snippets from various sources and prompt the model to generate programming problems inspired by these examples. This allows us to tap into a wide range of topics and create a comprehensive set of problem descriptions (as demonstrated in Figure~\ref{fig:embedding-plot}).


It is noted that code generation tasks can follow different problem description formats despite having the same content. For example, here is the same problem but stated in different ways:
\begin{itemize}[nosep,leftmargin=*]
    \item Write a python function to remove the kth element from a given list.
    \item In the ancient Library of Alexandria, scrolls are stored in a mystical list. The librarian needs to remove a specific scroll whenever a visitor requests it.
\end{itemize}

To make our prompt sets accommodate these diverse problem description formats, we adopt some templates from the training set of different coding benchmarks (e.g., MBPP, APPS) into the original prompt to generate the problem description.

To create a self-contained coding problem description, we need to ensure it not only includes a clear problem statement but also a well-defined starter code. The starter code should contain all necessary built-in libraries and a detailed function signature description to show what should be the input and output. Therefore, after obtaining the original problem description generated by the model, we then ask the model to generate function signatures. After deduplication, we get 103,280 problem descriptions in total.

\textbf{Code Generation}\ \ \ We prompt the LLM to solve each problem given the generated function signature. Following the Llama 3.1~\cite{dubey2024llama}, we also require the model to explain its thought process in comments, which improves code generation in both accuracy and readability.

\textbf{Test Generation}\ \ \ 
\label{sec:test-generation}
We use a pipelined approach to generate unit test sets. Specifically, we first ask the model to generate a set of valid inputs and then ask it to generate expected outputs based on the inputs.

For input generation, we ask the model to generate different types of function inputs to cover different cases including general, corner or difficult cases. For example, the following problem description should contain two different cases:

\fbox{%
  \parbox{\columnwidth}{%
\textbf{Problem Description:}

Write a function to find the longest string in a list of strings. If the strings are not comparable due to being of different lengths, the function should return None.

{\tt longest\_string(strings: list[str])}

\textbf{Case 1:} {\tt strings} is a list of strings.

\textbf{Case 2:} {\tt strings} is empty.

  }%
}

For output generation, we ask the model to generate the expected output based on the problem description and input. We also find that two strategies can boost the performance of unit test generation (as demonstrated in Section~\ref{sec:base-performance}):
\begin{itemize}[nosep, leftmargin=*]
    \item \textbf{Majority Voting}: The majority voting mechanism in the self-consistency approach~\cite{wang2022self} asks the model to generate multiple candidate outputs for a query, and aggregates them using a majority voting procedure to select the most commonly occurring answer.
    \item \textbf{Chain-of-Thought reasoning:} Following \citet{wei2022chain}, before outputting the expected values, we asked the model to first generate its reasoning steps.
\end{itemize}

Additionally, we employ the following strategies to ensure the quality and robustness of the generated unit tests:

\begin{itemize}[nosep, leftmargin=*]
    \item \textbf{Test Coverage Optimization:} We sample multiple candidate test cases and strategically select a subset that maximizes branch coverage of the solution code (a maximum coverage problem), ensuring comprehensive testing of different execution paths.
    \item \textbf{Output Diversification:} We notice that if the outputs of the test cases are not diverse enough, the solution code can cheat by exploiting patterns in test cases. For example, if all test cases return the same value (e.g., {\tt True}), the model could trivially pass by implementing a function that always returns that value. To address this issue, we explicitly select test samples with diverse output values.
\end{itemize}

\textbf{Synthetic Preference Data Generation}\ \ \ 
The key goal of constructing synthetic data for preference tuning, which is DPO training in our case, is to construct pairs of ``chosen'' and ``rejected'' responses to the given prompts. As illustrated in Section~\ref{sec:dpo-training}, the ``chosen'' examples are those examples where the solver  agrees with the verifier, i.e., the generated codes can pass the generated tests. However, identifying ``rejected'' examples is considerably more complex, as it is not always clear which side is at fault when they disagree.

In previous work, \citet{dong2024self} adopts an automated quality cross verification process by selecting both test case and functions with an accuracy rate grater than 0.5.
We utilize a similar cross validation strategy, but with a focus on both code and test generation. Specifically, if at least one generated solution passes all the generated tests, we will assume the solution and the test are ``correct''.
For training the solver, any other sampled code solutions that fail to pass these ``correct'' tests are treated as ``rejected'' examples.
Similarly, for training the verifier, when we find a ``correct'' test $f(x) == y$, we revisit the original sampling space for generating expected outputs, and treat any expected outputs $y' \neq y$ as ``rejected'' tests $f(x) == y'$. In this way, we can reuse all the chain-of-thought explanations generated by the model  during the majority voting process.

%% file: sections/04.experiments.tex
\section{Experiments}
\begin{table*}[ht]
\caption{Evaluation results over training iterations for our Solver-Verifier ( {\sc Sol-Ver}) method. $\Delta$(\%) means the relative percentage change from Baseline to Iter3$_\text{+DPO}$. Pass\% means average pass rate, Acc\% is accuracy, and FP\% is false positive rate.  {\sc Sol-Ver} provides large gains over the baseline on both code generation (solver) and unit test generation (verifier) tasks, which increase across iterations.}
\begin{center}
\resizebox{\textwidth}{!}{
\begin{tabular}{ll|lllllll|r}
\toprule
 &
   &
  \textbf{Baseline} &
  \textbf{Iter1$_\text{+SFT}$} &
  \textbf{Iter1$_\text{+DPO}$} &
  \textbf{Iter2$_\text{+SFT}$} &
  \textbf{Iter2$_\text{+DPO}$} &
  \multicolumn{1}{c}{\textbf{Iter3$_\text{+SFT}$}} &
  \multicolumn{1}{c|}{\textbf{Iter3$_\text{+DPO}$}} &
  \textbf{$\Delta$(\%)} \\
\midrule
\multirow{2}{*}{\textbf{MBPP}}          & Code$_\text{Pass\%}$ & 38.60 & 38.60 & 40.80 & 38.60 & 40.80 & 40.80 & 41.00 & 6.17  \\
                                        & Test$_\text{Acc\%}$ & 42.68 & 49.01 & 49.50 & 50.69 & 51.01 & 51.54 & 51.76 & 17.54 \\
                                        & Test$_\text{FP\%}$ & 12.75 & 12.57 & 10.32 & 10.12 & 10.06 & \ \ 9.80 & \ \ 9.60 & 24.71 \\
\midrule
\multirow{2}{*}{\textbf{LiveCodeBench}} & Code$_\text{Pass\%}$ & 18.23 & 24.86 & 25.97 & 25.12 & 26.38 & 26.41 & 27.24 & 33.08 \\
                                        & Test$_\text{Acc\%}$ & 20.14 & 36.43 & 37.09 & 39.40 & 40.65 & 41.25 & 41.50 & 51.47 \\
                                        & Test$_\text{FP\%}$ & 20.76 & 20.65 & 19.34 & 19.28 & 19.05 & 18.94 & 18.63 & 10.26 \\
\bottomrule
\end{tabular}}
\end{center}
\label{tab:full-performance}
\end{table*}

\subsection{Experimental Setup}

\textbf{Models and Datasets}\ \ \ We conduct experiments using Llama 3.1 8B~\cite{dubey2024llama}.\footnote{\url{https://huggingface.co/meta-llama/Llama-3.1-8B}} For the SFT and DPO training, we use the fairseq2 infrastructure~\cite{balioglu2023fairseq2} and run inference using vLLM~\cite{kwon2023efficient}.

We sample the problem descriptions using Llama 3.1 based on open source snippets from the OSS-Instruct dataset~\cite{wei2024magicoder}\footnote{\url{https://huggingface.co/datasets/ise-uiuc/Magicoder-OSS-Instruct-75K}}. For the problem description templates, we use the training sets of some  standard coding benchmark: MBPP~\cite{austin2021program}, APPS~\cite{hendrycksapps2021} and CodeContest~\cite{li2022competition}. For our experoiments we only focus on Python-related coding questions.

We test both code generation and unit test generation on standard python coding benchmarks including:
\begin{itemize}[nosep,leftmargin=*]
    \item MBPP~\cite{austin2021program}: A popular benchmark for Python code generation which focuses on relatively simple, self-contained functions.
    \item LiveCodeBench~\cite{jain2024livecodebench}: A comprehensive and contamination-free evaluation of LLMs for code, which continuously collects new problems over time from contests across three competition platforms, namely LeetCode, AtCoder, and CodeForces.
\end{itemize}

LiveCodeBench contains both code generation and test output prediction tasks. Since MBPP does not originally include a unit test generation task, we utilize the existing gold unit tests to assess the model’s accuracy in generating expected outputs given the inputs in gold unit tests.

\textbf{Evaluation Metrics}\ \ \ For the code generation task, all results are obtained using greedy decoding with the pass@1 metric (Pass\%). For unit test generation, we consider the following metrics:
\begin{itemize}[nosep,leftmargin=*]
    \item Accuracy (Acc\%): The accuracy of test output prediction. A test output is correct only when its literal value is equal to the gold one.
    \item False Positive Rate (FP\%): A robust unit test set should effectively differentiate between correct and incorrect code solutions. To evaluate this capability, in addition to the pass rate on correct solutions, we also evaluate the pass rate for known flawed solutions (false positive rate). To obtain flawed solutions for testing, we generate 20 candidate code solutions for each coding problem and use the gold unit tests to identify those that fail, treating them as negative examples.\footnote{We use the same Llama 3.1 8B base model to generate negative examples (temperature is set to 0.6 and top p is set to 0.9). As a result, we get 400 examples for both MBPP and LiveCodeBench separately.} 
\end{itemize}

\subsection{Evaluating {\sc Sol-Ver}}
Table~\ref{tab:full-performance} reports  {\sc Sol-Ver}'s iterative training performance for both code and test generation tasks on the MBPP and LiveCodeBench datasets.\footnote{Our baseline performance for code generation differs from that reported in the Llama 3.1 technical report because we employ a unified three-shot prompt template for both MBPP and LiveCodeBench, rather than using the prompts specifically provided for MBPP. We do not include the gold test in the prompt.}
To further show the improvement trend for {\sc Sol-Ver}, we plot the performance change across iterations for test generation in Figure~\ref{fig:test-gen-perf} (other figures for iterative change can be found in Appendix~\ref{app:performance-change}).
The results demonstrate that {\sc Sol-Ver} can consistently improve the performance of the base Llama 3.1 model in both code generation and unit test generation, as illustrated by the increased pass rate for code generation, increased accuracy and decreased false positive rate for test generation at each iteration. Specifically, we have achieved an average of 19.63\% and 17.49\% relative improvement for code and test generation respectively. In particular, two interesting conclusions can be made:
\begin{itemize}[nosep, leftmargin=*]
    \item The improvement in unit test generation performance is more significant than that observed in code generation. This greater enhancement may be due to the relatively limited availability of unit test–related code data compared to code generation–related data during the pre-training phase of Llama 3.1.
    \item The difference between SFT+DPO and SFT-only models suggests that incorporating negative examples during preference tuning helps the model learn from errors and refine its generation strategies.
\end{itemize}

\begin{figure}
    \centering
    \includegraphics[width=0.8\linewidth]{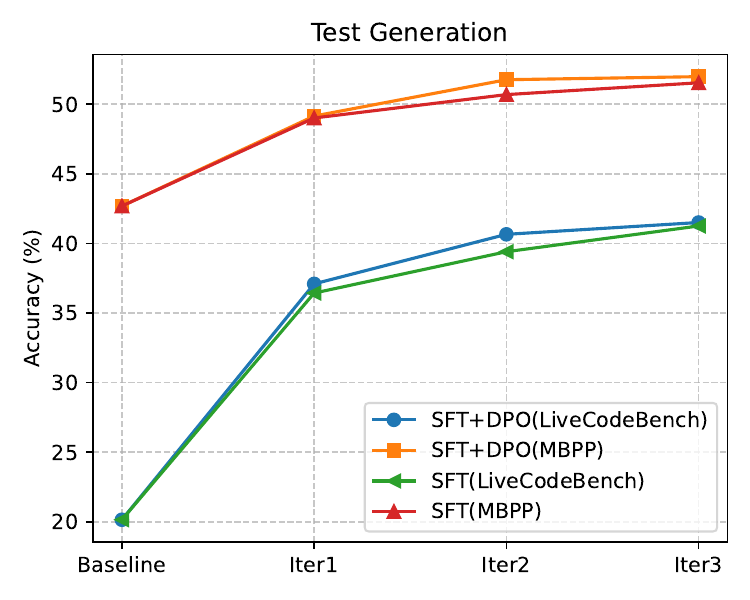}
    \vspace{-0.5em}
    \caption{Iterative performance of our method, {\sc Sol-Ver}, for test generation. Our method outperforms the baseline and SFT training for both LiveCodeBench and MBPP benchmarks, and improve across training iterations.  }
    \label{fig:test-gen-perf}
\end{figure}

\subsection{Evaluating the Base Model}
\label{sec:base-performance}
In Section~\ref{sec:practical-settings}, we discuss several strategies for improving test generation. To evaluate their effectiveness and assess the initial performance of both code generation (LLM-as-a-solver) and test generation (LLM-as-a-verifier), we conduct base model evaluation using Llama 3.1 8B on the MBPP benchmark. Specifically, we measure: (1) the pass rate of generated code as evaluated by gold standard unit tests; and (2) the pass rate of generated unit tests when executed against the gold standard code solutions. The results are presented in the first two rows of Table~\ref{tab:base-model-eval}.

\begin{table}[t]
\caption{Code generation and test generation performance for Llama 3.1 8B on the MBPP benchmark. The case pass rate represents the average pass rate per test set. CoT means Chain-of-Thought reasoning, and MV means majority voting.}
\vspace{-0.5em}
\begin{center}
\resizebox{\columnwidth}{!}{
\begin{tabular}{l|cc}
\toprule
\textbf{Llama 3.1 8B}               & \textbf{Pass Rate} & \textbf{Case Pass Rate} \\
\midrule
Code Generation        & 38.60\%            & 44.20\%                      \\
Test Generation & 18.60\%            & 37.60\%\\
\ \ + CoT & 19.60\% & 39.60\%\\
\ \ + MV & 24.80\%  & 42.40\% \\
\ \ + CoT + MV & 31.40\% & 47.60\%\\
\midrule
Code Reranked by Synthetic Test Rate & 35.00\% & 42.40\%\\
\bottomrule
\end{tabular}}
\end{center}
\label{tab:base-model-eval}
\end{table}


As shown in the table, unit test generation significantly underperforms compared to code generation on the same set of examples. This highlights a performance gap between employing LLMs as solvers and as verifiers. This discrepancy motivates our work, as we aim to enhance verification capabilities by leveraging the strong code generation abilities of LLMs. Additionally, improvements in unit test generation can, in turn, contribute to stronger code generation.

To further improve the baseline performance on unit test generation, we utilize two sampling strategies as discussed in Section~\ref{sec:test-generation}, i.e., Majority Voting (MV) and Chain-of-Thought (CoT) Prompting, that can improve the accuracy of generated unit tests. These strategies were originally proven effective for other tasks, such as mathematical reasoning. In Table~\ref{tab:base-model-eval}, we report the results after applying these strategies. As can be seen, both strategies can significantly improve the quality of unit test generation on the MBPP datasets, and combining them together yields the best result.

Nevertheless, despite these improvements, the combined strategies remain insufficient for generating an optimal test set for verification.
To substantiate this claim, we report the performance of code generation reranked by the generated tests (see the last row of Table~\ref{tab:base-model-eval}). Specifically, multiple code solution candidates are sampled and ranked using the generated tests as outlined in ``Test Generation + CoT + MV'' from Table~\ref{tab:base-model-eval}. The results indicate a decline in performance compared to the original code generation, thereby demonstrating that poor-quality test generation adversely affects the final outcomes. In contrast, our proposed method, {\sc Sol-Ver}, effectively addresses this limitation.

\subsection{Evaluating Agreement between Iterations}

Since models from different iterations are trained on different sets of synthetic data, which may introduce varying biases, we examine the agreement between the first and second iteration models to monitor progress across iterations. Specifically, both models are employed to generate unit tests for the MBPP benchmarks, and we measure the extent to which the unit tests produced by each model can agree on the gold code solutions. The results are given in Table~\ref{tab:agreement}.
Both datasets exhibit performance enhancements from the first to the second iteration and maintain a high level of agreement across iterations. This indicates that, despite being trained on distinct sets of synthetic data, both iterations produce largely consistent unit test generation outputs.

\begin{table}[t]
\caption{Agreement between Iter 1 and Iter 2, and the test accuracy for the model ensemble. The ensemble approach (Ens.) refers to selecting code solutions that successfully pass both the tests generated in the first iteration and those from the second iteration.}
\label{tab:agreement}
\begin{center}
\resizebox{\columnwidth}{!}{
\begin{tabular}{l|lll|l}
\toprule
\textbf{Dataset}       & \textbf{Acc\%$_\text{Iter1}$}   & \textbf{Acc\%$_\text{Iter2}$}   & \textbf{Agreemnt} & \textbf{Acc\%$_\text{Ens.}$} \\
\midrule
MBPP          & 49.50\% & 51.01\% & 75.14\%  & 51.12\%   \\
LiveCodeBench & 37.09\% & 40.65\% & 72.38\%  & 40.68\%  \\
\bottomrule
\end{tabular}}
\end{center}
\end{table}

Given the agreement between iterations, we evaluate the performance of a model ensemble.
In this context, the ensemble approach refers to selecting code solutions that successfully pass both the tests generated in the first iteration and those from the second iteration.
The ensemble demonstrated a modest improvement over the second iteration alone, indicating that integrating the outputs from both iterations can enhance overall performance. As a result, we adopt this ensemble method when generating synthetic data for the third iteration for {\sc Sol-Ver}, as reported in Table~\ref{tab:full-performance}.

\begin{table*}[h]
\caption{Illustrative examples of test case refinement by {\sc Sol-Ver}'s verifier component across iterations for two distinct programming problems. \colorbox{pink}{Red highlighting} indicates tests that were incorrect or suboptimal in earlier iterations but were corrected or improved by later iterations, demonstrating the verifier's learning progress.}
\begin{center}
\resizebox{\textwidth}{!}{
\begin{tabular}{l|l|l}
\toprule
Problem &
  \begin{tabular}[c]{@{}l@{}}Write a function that takes an integer number of seconds\\as input and returns the number of minutes in that time,\\disregarding any remaining seconds.\end{tabular} &
  \begin{tabular}[c]{@{}l@{}}Write a function that calculates and returns the greatest\\common divisor (GCD) of two integers using the Euclidean\\algorithm, with the output formatted as "GCD(a, b)= result".\end{tabular} \\\midrule
Iter 1 &
  \begin{tabular}[c]{@{}l@{}}\texttt{assert minutes\_in(1) == 0}\\ \texttt{\colorbox{pink}{assert minutes\_in(3660 + 60 + 60 + 1) == 3}}\\ \texttt{\colorbox{pink}{assert minutes\_in(3660 + 60 + 60 + 60 + 60 + 1) == 5}}\end{tabular} &
  \begin{tabular}[c]{@{}l@{}}\texttt{\colorbox{pink}{assert pgcd(8, 7) == 1}}\\ \texttt{\colorbox{pink}{assert pgcd(20, 25) == 5}}\\ \texttt{\colorbox{pink}{assert pgcd(14, 6) == 2}}\end{tabular} \\
\midrule
Iter2 &
  \begin{tabular}[c]{@{}l@{}}\texttt{assert minutes\_in(1) == 0}\\ \texttt{\colorbox{pink}{assert minutes\_in(3660 + 60 + 60 + 1) == 61}}\\ \texttt{\colorbox{pink}{assert minutes\_in(3660 + 60 + 60 + 60 + 60 + 1) == 61}}\end{tabular} &
  \begin{tabular}[c]{@{}l@{}}\texttt{\colorbox{pink}{assert pgcd(8, 7) == "PGCD(8,7) = 1"}}\\ \texttt{\colorbox{pink}{assert pgcd(20, 25) == "PGCD(20,25) = 5"}}\\ \texttt{\colorbox{pink}{assert pgcd(14, 6) == "PGCD(14,6) = 2"}}\end{tabular} \\
\midrule
Iter3 &
  \begin{tabular}[c]{@{}l@{}}\texttt{assert minutes\_in(1) == 0}\\ \texttt{assert minutes\_in(3660 + 60 + 60 + 1) == 63}\\ \texttt{assert minutes\_in(3660 + 60 + 60 + 60 + 60 + 1) == 65}\end{tabular} &
  \begin{tabular}[c]{@{}l@{}}\texttt{assert pgcd(8, 7) == "GCD(8, 7) = 1"}\\ \texttt{assert pgcd(20, 25) == "GCD(20, 25) = 5"}\\ \texttt{assert pgcd(14, 6) == "GCD(14, 6) = 2"}\end{tabular} \\
\bottomrule
\end{tabular}}
\end{center}
\label{tab:case-study}
\end{table*}

%% file: sections/05.ablation-study.tex
\section{Ablation Study}
\subsection{Prompt and Test Analysis}
\begin{figure}[t]
    \centering
    \includegraphics[width=0.9\columnwidth]{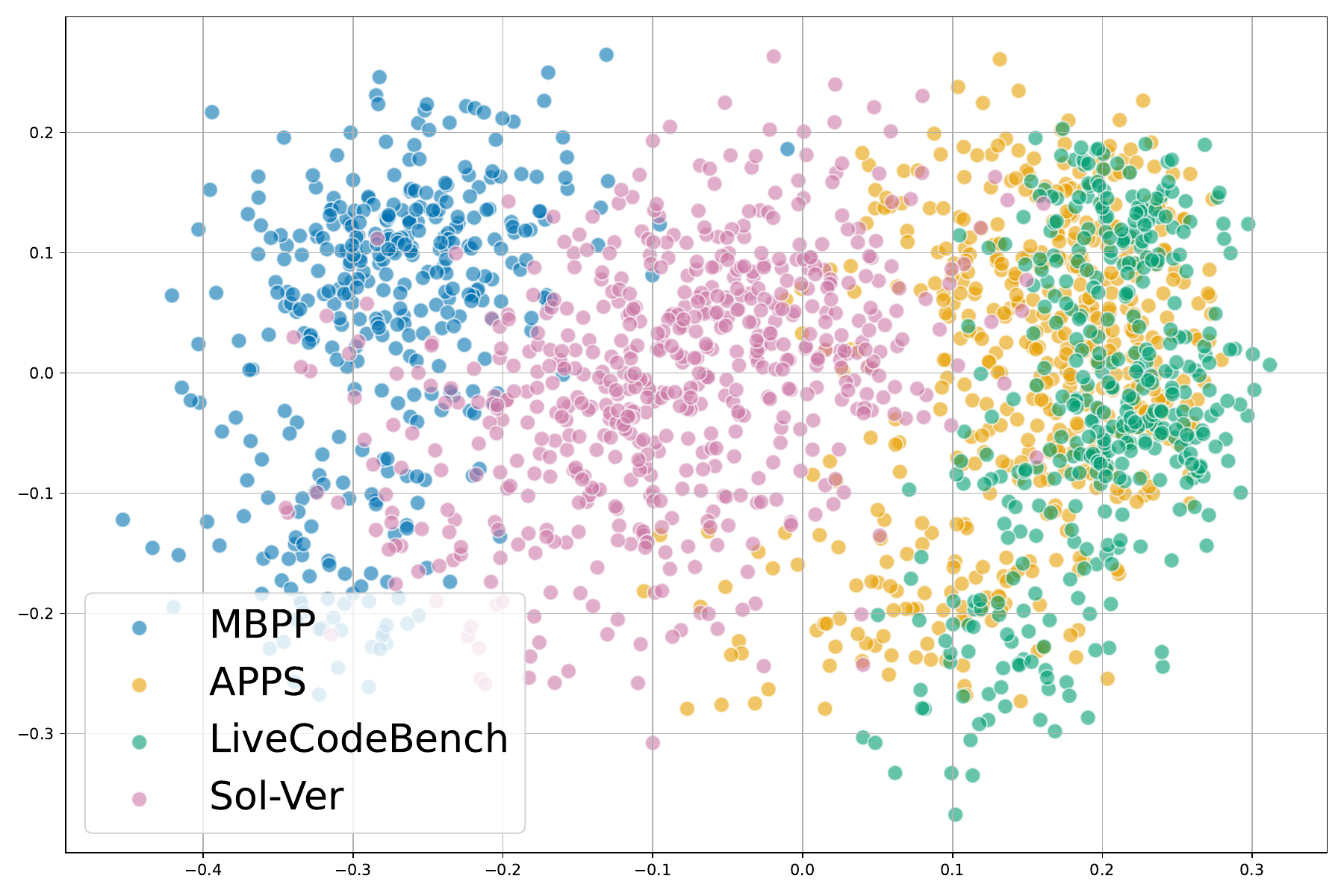}
    \caption{Prompt distribution comparison with other standard coding benchmarks. We use Principal Component Analysis (PCA) for embedding dimension reduction.}
    \label{fig:embedding-plot}
\end{figure}

\textbf{Prompt Coverage Analysis}\ \ \ To analyze the domain coverage of our generated coding problem set, in Figure~\ref{fig:embedding-plot}, we visualize the embedding distributions of our synthetic problem descriptions with those from established coding benchmarks, including MBPP, APPS, and LiveCodeBench.
Specifically, we use Gecko, a compact and versatile text embedding model distilled from LLMs~\cite{lee2024gecko} for obtaining the sentence embeddings.
The embedding distribution plot reveals that our synthetic problem set exhibits a broad and diverse coverage, encompassing a wide range of topics, difficulty levels, and programming paradigms present in the compared benchmarks.

\begin{figure}
    \centering
    \includegraphics[width=\columnwidth]{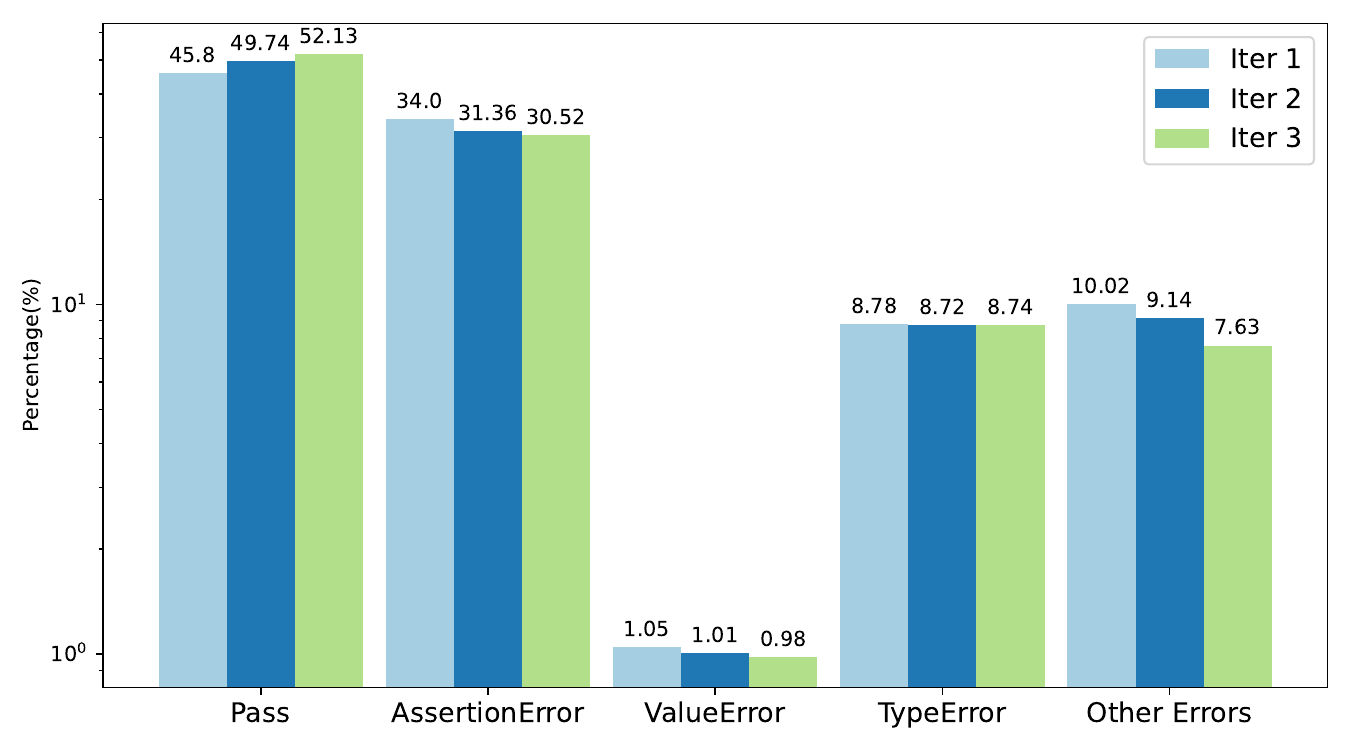}
    \caption{Pass or error distribution of synthetic data generated at each iteration.}
    \label{fig:error-dist}
\end{figure}

\textbf{Progress Analysis per Iteration}\ \ \ To evaluate the iterative advancements of our model, we present a case study on test generation across iterations in Table~\ref{tab:case-study}. The results illustrate that \textsc{Sol-Ver} progressively refines its test generation for the same set of coding problems, thereby enhancing the quality of the synthetic data. These enhancements include the generation of more accurate expected values and better adherence to required format specifications.
Additionally, we monitor the execution results at each iteration and display the distribution of pass and error rates in Figure~\ref{fig:error-dist}. As shown, the pass rate increases with each iteration, primarily due to a reduction in assertion errors, indicating an improvement in the accuracy of the predicted expected outputs.

\begin{table}[h]
\caption{Code generation performance for different settings of the scoring function for selecting DPO pairs. Results indicate better results are obtained with less, but higher quality, data ($\epsilon>0$).}
\begin{center}
\resizebox{\columnwidth}{!}{
\begin{tabular}{l|cccc}
\toprule
$\epsilon$    & \multicolumn{1}{c}{$> 0$} & \multicolumn{1}{c}{$> 0.5$} & \multicolumn{1}{c}{$ > 0.75$} & \multicolumn{1}{c}{\textsc{Sol-Ver}}\\
\midrule
Data Size &25,525 & 20,457 & 13,158 & 12,525\\
MBPP          & \multicolumn{1}{c}{36.00} & \multicolumn{1}{c}{37.00}   & \multicolumn{1}{c}{\textbf{40.80}}  &   \multicolumn{1}{c}{\textbf{40.80}} \\
LiveCodeBench & \multicolumn{1}{c}{22.41} & \multicolumn{1}{c}{25.75}   & \multicolumn{1}{c}{\textbf{26.18}}   & \multicolumn{1}{c}{25.97}  \\
\hline \hline
$\text{Score}(C^-, T)$ & Random & Lowest & Median & \textsc{Sol-Ver}\\
\midrule
MBPP                   & \textbf{40.80}  & 39.00  & 38.60  & \textbf{40.80}\\
LiveCodeBench          & \textbf{26.18}  & \textbf{26.18}  & 25.98  & 25.97\\
\bottomrule
\end{tabular}}
\end{center}
\label{tab:scoring-func}
\end{table}

\subsection{Discussion on the Scoring Function}
In Section~\ref{sec:problem-formulation}, we define the scoring function for selecting the chosen / rejected solution-test pair as a binary function.
In our experiments for Iter 1, we find that only 45\% examples can get agreed solution-test pairs, where there is at least one code generated by the model will pass all the generated test. Consequently, the total number of preference tuning pairs is limited by the number of these selected examples.

To explore whether we can utilize the rest of the data where the pass rate is not necessarily 100\%, but is still high enough to rely on, we conduct a series of experiments to discuss the potential of using a \textit{soft pass rate} for selecting preference pairs. Specifically, we change the chosen / rejected pair as $(C^-, C^+)$, where for the same test suite $\mathbf{T}$, $\text{Score}(C^+, \mathbf{T}) > \text{Score}(C^-, \mathbf{T})$, and $\text{Score}(C^+, \mathbf{T}) \geq \epsilon$, where $\epsilon$ is a threshold to determine above which pass rate the test set is relatively reliable. For simplicity, we discuss three cases for $\epsilon$: (1) $\epsilon > 0$ (can be any number); (2) $\epsilon > 0.5$; (3) $\epsilon > 0.75$. For $\text{Score}(C^-, \mathbf{T})$, we also consider three cases: (1) $\text{Score}(C^-, \mathbf{T})$ is a random score; (2) $\text{Score}(C^-, \mathbf{T})$ is the lowest score among all sampling candidate; (3) $\text{Score}(C^-, \mathbf{T})$ is the median score from the lowest to $\text{Score}(C^+, \mathbf{T})$.

In Table~\ref{tab:scoring-func}, we first present various $\epsilon$ settings for assigning $\text{Score}(C^-, \mathbf{T})$ as a random score. After identifying the optimal setting from our results ($\epsilon > 0.75$), we examine the impact on $\text{Score}(C^-, \mathbf{T})$. The findings reveal that randomly selecting the threshold for a high-quality test set significantly degrades performance, despite an increase in data size. This underscores that the quality of synthetic data is more critical than its quantity. Regarding $\text{Score}(C^-, \mathbf{T})$, we found that its impact is less sensitive compared to $\epsilon$. Considering the overall performance, we have chosen to retain the original settings for \textsc{Sol-Ver} to maintain simplicity. We also encourage future research to explore the potential of scoring functions design in more diverse ways.

%% file: sections/06.conclusion.tex
\section{Conclusion}

In this work, we introduced {\sc Sol-Ver}, a novel self-play framework designed to address the critical challenge of data scarcity and the performance disparity between LLM-based code solvers and test verifiers. By enabling an LLM to simultaneously embody both roles, {\sc Sol-Ver} facilitates a co-evolutionary process where improvements in test generation lead to better code synthesis, and vice-versa. This iterative refinement loop generates high-quality synthetic code-test pairs, demonstrably enhancing both capabilities without reliance on human annotations or larger teacher models. Our experiments with Llama 3.1 8B validate {\sc Sol-Ver}'s efficacy, achieving significant performance gains on established benchmarks. {\sc Sol-Ver} represents a step towards more autonomous, data-efficient, and robust systems for automated code and test generation, offering a scalable and adaptable paradigm for continuous self-improvement in code AI.

There are several avenues for further enhancement of {\sc Sol-Ver}. Future work could explore more controlled input generation techniques to optimize the relevance and quality of synthetic data. Additionally, extending the framework to accommodate more complex coding scenarios beyond function-level generation can broaden its applicability.
Further optimization of the iterative training process could also reduce computational overhead, making the approach more efficient.
Finally, evaluating the framework on larger and more diverse models will help to determine its generalizability and effectiveness across different architectures.

%% file: sections/appendix.tex
\section{Prompt Template}
\label{app:prompt-template}
\begin{tcolorbox}[
    colback=gray!10,           
    colframe=gray!75,          
    title=Prompt for Generating Problem Description,      
    fonttitle=\bfseries,       
    width=\textwidth,          
    sharp corners,             
    boxrule=0.75mm,            
    coltitle=black             
]
[Code Snippet]

{\tt for (first\_p, second\_p) in zip\_longest(diag1, diag2):}

{\tt\ \ \ \ assert first\_p[0] == pytest.approx(second\_p[0])}

{\tt\ \ \ \ assert first\_p[1] == pytest.approx(second\_p[1])}

\ 

[Template]

Polycarp is reading a book consisting of $n$ pages numbered from $1$ to $n$. Every time he finishes the page with the number divisible by $m$, he writes down the last digit of this page number. For example, if $n=15$ and $m=5$, pages divisible by $m$ are $5, 10, 15$. Their last digits are $5, 0, 5$ correspondingly, their sum is $10$.

Your task is to calculate the sum of all digits Polycarp has written down.

You have to answer $q$ independent queries.

\ 

[Instruction]

Please gain inspiration from the previous random code snippet and template to create a high-quality python programming problem.

Rules:

- Never mention the ``code snippet''.

- Don't write the solution.

- Do not specify constraints nor example inputs/outputs.

- The inspiration is just an inspiration. You can deviate from it.

- The solution of the problem should be only one function, not an entire program.

- The problem should be self-contained.
\end{tcolorbox}

\begin{tcolorbox}[
    colback=gray!10,           
    colframe=gray!75,          
    title=Prompt for Generating Function Signature,      
    fonttitle=\bfseries,       
    width=\textwidth,          
    sharp corners,             
    boxrule=0.75mm,            
    coltitle=black             
]
\textless Problem1\textgreater

Write a function to find the similar elements from the given two tuple lists.

\textless /Problem1\textgreater

\textless Signature1\textgreater

similar\_elements(test\_tup1: list, test\_tup2: list) -\textgreater list

\textless/Signature1\textgreater

\textless Problem2\textgreater

Write a python function to identify non-prime numbers.

\textless/Problem2\textgreater

\textless Signature2\textgreater
\end{tcolorbox}

\begin{tcolorbox}[
    colback=gray!10,           
    colframe=gray!75,          
    title=Prompt for Generating Test Input,      
    fonttitle=\bfseries,       
    width=\textwidth,          
    sharp corners,             
    boxrule=0.75mm,            
    coltitle=black             
]
\textless Q1\textgreater

Write a function to find the longest string in a list of strings. If the strings are not comparable (due to being of different lengths), the function should return None.
function signature: longest\_string(strings: list[str])

\textless/Q1\textgreater

\textless ANALYSIS1\textgreater

- Case 1: `strings` is a list of strings.

- Case 2: `strings` is empty.

\textless/ANALYSIS1\textgreater

\textless INPUTS1\textgreater

{\tt longest\_string(['dog', 'cat', 'elephant']) \# Consider case 1.}

{\tt longest\_string([]) \# Consider case 2.}

\textless/INPUTS1\textgreater

\textless Q2\textgreater

Write a function to check if the given string represents a sequence of ASCII characters.
The function should be able to handle different types of sequences, such as lists, tuples, and NumPy arrays.
The function should return True if the sequence contains only ASCII characters, and False otherwise.

function signature: is\_ascii(seq: list) -\textgreater bool

\textless/Q2\textgreater

\textless ANALYSIS2\textgreater

\end{tcolorbox}

\begin{tcolorbox}[
    colback=gray!10,           
    colframe=gray!75,          
    title=Prompt for Generating Test Output,      
    fonttitle=\bfseries,       
    width=\textwidth,          
    sharp corners,             
    boxrule=0.75mm,            
    coltitle=black             
]
\textless Q1 \textgreater

Write a function to find the similar elements from the given two tuple lists.

function signature: {\tt similar\_elements(test\_tup1: Tuple, test\_tup2: Tuple)}

\textless/Q1 \textgreater

\textless INPUT1 \textgreater

{\tt similar\_elements((3, 4, 5, 6),(5, 7, 4, 10))}

\textless /INPUT1 \textgreater

\textless ANALYSIS1 \textgreater

- In the first tuple (3, 4, 5, 6), the elements 4 and 5 are present.

- In the second tuple (5, 7, 4, 10), the elements 4 and 5 are also present.

- Since 4 and 5 are present in both tuples, they should be included in the output.

- The other elements in the tuples (3, 6, 7, and 10) are not present in both tuples, so they should not be included in the output.

- So the expected output is (4, 5).

\textless /ANALYSIS1 \textgreater

\textless OUTPUT1 \textgreater

{\tt assert similar\_elements((3, 4, 5, 6),(5, 7, 4, 10)) == (4, 5)}

\textless /OUTPUT1 \textgreater

\textless Q2 \textgreater

Write a python function to identify non-prime numbers.

function signature: {\tt is\_not\_prime(n: int)}

\textless /Q2 \textgreater

\textless INPUT2 \textgreater

{\tt is\_not\_prime(2)}

\textless /INPUT2 \textgreater

\textless ANALYSIS2 \textgreater

- One of the fundamental properties of prime numbers is that they can only be divided evenly by 1 and themselves.

- 2 is considered a prime number because it can only be divided evenly by 1 and itself.

- So 2 is a prime, and the expected output is False.

\textless /ANALYSIS2 \textgreater

\textless OUTPUT2 \textgreater

{\tt assert is\_not\_prime(2) == False}

\textless /OUTPUT2 \textgreater

\textless Q3 \textgreater

Write a function to find all words which are at least 4 characters long in a string by using regex.

function signature: {\tt find\_char\_long(text: str)}

\textless /Q3 \textgreater

\textless INPUT3 \textgreater

{\tt find\_char\_long('Jing Eco and Tech')}

\textless /INPUT3 \textgreater

\textless ANALYSIS3 \textgreater

- For the first word 'Jing', it's 4 characters long, and 4 >= 4, so it should be included.

- For the second word 'Eco', it's 3 characters long, and 3 < 4, so it should NOT be 
included.

- For the third word 'and', it's 3 characters long, and 3 < 4, so it should NOT be included.

- For the fourth word 'Tech', it's 4 characters long, and 4 >= 4, so it should be included.

- To sum up, the output is ['Jing', 'Tech'].

\textless /ANALYSIS3 \textgreater

\textless OUTPUT3 \textgreater

{\tt assert find\_char\_long('Jing Eco and Tech') == ['Jing', 'Tech']}

\textless /OUTPUT3 \textgreater

\textless Q4 \textgreater

{}

function signature: {}

\textless /Q4 \textgreater

\textless INPUT4 \textgreater

{}

\textless /INPUT4 \textgreater

\textless ANALYSIS4 \textgreater
\end{tcolorbox}

\section{Performance Change Across Iterations}
\label{app:performance-change}
\begin{figure*}[ht]
    \centering
    \begin{subfigure}[b]{0.4\textwidth}
        \centering
        \includegraphics[width=\textwidth]{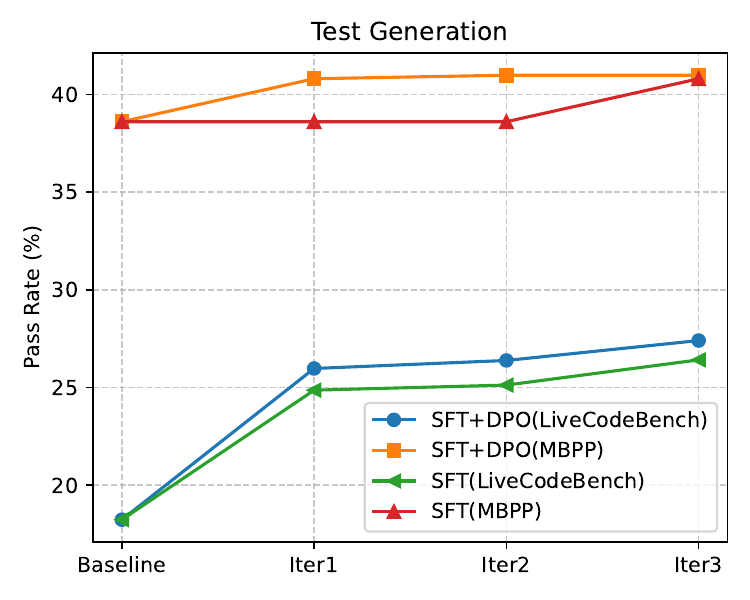}
    \end{subfigure}
    \begin{subfigure}[b]{0.4\textwidth}
        \centering
        \includegraphics[width=\textwidth]{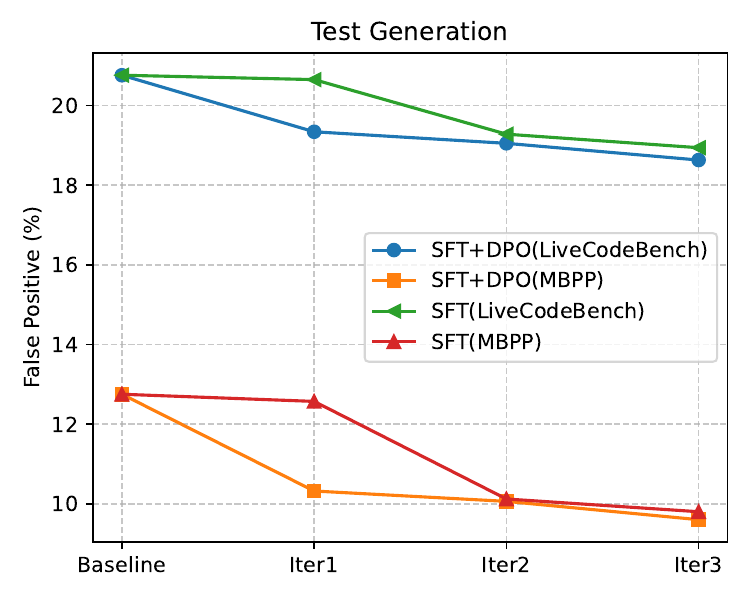}
    \end{subfigure}
    \caption{Performance of Solver-Verifier Framework.}
    \label{fig:solver-verifier-perf}
\end{figure*}